\documentclass[prl,showpacs,twocolumn]{revtex4}
\usepackage[usenames]{color}
\usepackage{bm}
\usepackage{amsmath}
\usepackage{amssymb}
\usepackage{epsfig}

\newcommand{\be}{\begin{equation}}
\newcommand{\en}{\end{equation}}
\newcommand{\rd}{\mathrm{d}}

\begin{document}

\title{The quantum  de Finetti representation for the Bayesian  Quantum Tomography and the Quantum Discord }

\author{ V. S. Shchesnovich$^{1}$ and D. S. Mogilevtsev$^{1,2}$  }

\affiliation{${}^1$Centro de Ci\^encias Naturais e Humanas, Universidade Federal do
ABC, Santo Andr\'e,  SP, 09210-170 Brazil\\ ${}^2$Institute of Physics, Belarus
National Academy of Sciences, F.Skarina Ave. 68, Minsk 220072 Belarus}

\begin{abstract}

We point out that  the quantum de Finetti representation, unique for infinitely extendable exchangeable systems, 
assigns a non-zero Quantum Discord  to (quantumly) uncorrelated  systems  and thus  cannot serve as an universal prior distribution in
the Bayesian Quantum Tomography. This apparent paradox stems from  linearity of the Born rule for  the probability assignment 
in   Quantum Mechanics, which results in mixing of one's knowledge about the quantum state and the representative of the state 
in one density matrix.   
\end{abstract}
 \maketitle

Quantum Mechanics is formulated  with the concept of quantum state, a density matrix $\rho(t)$ in general, which  gives  the  probabilities  of  outcomes of all possible  measurements on the system. The   probabilities  are  defined  by the Positive Operator Valued Measure (POVM): $\Pi_\alpha\ge0$, $\alpha = 1,\ldots ,K$, $\sum_\alpha \Pi_\alpha = I$ with  the probability assignment being linear in the density matrix: $p_\alpha = \mathrm{Tr}(\Pi_\alpha \rho)$, which is the well-known result of Gleason's theorem \cite{Gleason} (see also Ref.~\cite{Busch}).
By measuring the quorum of observables one can obtain the
set of data sufficient for complete characterization of the density matrix. Feasibility of
such a procedure was brilliantly demonstrated in the beginning of 1990-es
\cite{VogelRisken,DAriano,Smithey, Durra} (actually, the firstly
suggested quantum homodyne tomography  scheme was proved to be such
a powerful and efficient tool, that the whole field of quantum
states/processes reconstruction was aptly nicknamed as
``Quantum Tomography"  (QT) \cite{all}). 

Due to irreversible  character and statistical  nature of  the quantum measurements, to perform the QT  the experimentalist needs
 \textit{an ensemble  of systems in identically prepared states.} Then, a statistical estimation procedure can be devised
allowing one to   estimate  the   density matrix parameters  (e. g. by using the  Maximal Likelihood Estimation
\cite{Hradil,Hradil_Mog}  or by resorting to the full Bayesian Statistical Inference  \cite{Jones,BayesQT,QBayes,QBayesD,BK}). If the number $N$  of systems  in the ensemble is sufficiently
large, the result of estimation is expected to converge to the actual density matrix. 

The quantum no-cloning theorem rules out duplicating of an unknown quantum state
\cite{nocloning}, thus the QT is always a process of  updating
of information about the state (if necessary,  updating    also the parameters 
of the  measurement device  \textit{at the  same time} \cite{mogilevtsev
calibration}).  Such an updating is the main idea behind  the 
Bayesian Statistical Inference method. The convergence property of
the likelihood function for a generic   measurement  on the ensemble (i.e. the
multinomial distribution) to the Dirac delta-function in the limit
of infinite number of measurements \cite{Gnedenko} (see also Ref.
\cite{BayesQT}) results in  agreement between different Bayesian experimentalists.
The Bayesian approach in the QT  was pioneered by K. R. W. Jones
\cite{Jones}, who obtained an upper bound on the accessible
information obtainable from measurement of a pure quantum state,
when the latter is represented by an invariant prior measure,  and
indicated  a measurement scheme (the so-called isotropic scheme)
which saturates this bound asymptotically.  The Bayesian approach
to the QT for the system consisting of $1/2$-spins  was
extensively studied in Ref. \cite{BayesQT}, where pure as well as
mixed quantum states of such  spin systems were considered.

We note that measurements in the QT are not restricted  to separate measurements 
on   individual systems of the ensemble.  For a finite $N$ it is more efficient to
measure the entire ensemble as a combined system, see Refs. \cite{Peres,Massar,Derka}. 
As single members of the ensemble are concerned, it is shown that asymptotically 
to the order of $1/N$ the effectiveness of the individual measurements approaches that
of the combined scheme \cite{Bagan}. However, the combined measurement on several 
systems of the ensemble allows one to uncover also the  correlations between the 
individual  ensemble members.

The central question in the Bayesian QT  is the way to represent one's 
incomplete  knowledge about an ensemble. Using  two  premises about the ensemble, namely that  of mutual exchangeability 
of the individual systems   and infinite extendability of the ensemble,  a unique   answer to this question is  the quantum de 
Finetti representation, which follows from the classical de Finetti theorem and Gleason's theorem  \cite{QdeFinetti,QBayesD}.  
It also  gives the  quantum Bayesian rule for  updating the probability distribution \cite{QBayes}.  The quantum de Finetti representation  
has the following    form 
\be 
\label{E1} \rho^{(N)} =
\int\rd\mu(\rho)\rho^{\otimes N}, \quad \rho^{\otimes N} \equiv
\underbrace{\rho\otimes\rho\otimes\ldots\otimes\rho}_{N}, 
\en
where  $\rd\mu(\rho) = \rd\rho P(\rho)$ and $P(\rho)$ is the
probability density. The Bayesian QT   then proceeds as follows. The
experimentalist's prior  knowledge about the state  is reflected in   the prior probability density
$P(\rho)$.  The Bayes rule for updating the probability density
after measurements on the first $M$ systems with the data $\Pi_1,
\ldots, \Pi_M$  (i.e. the POVM measurement  results for one or
several POVMs) reads
 \be
 \label{E2}
P(\rho|\Pi_1,\ldots,\Pi_M) = \frac{P(\Pi_1, \ldots, \Pi_M|\rho)P(\rho)}{P(\Pi_1,\ldots,\Pi_M)},
 \en
where \be \label{E3} P(\Pi_1,\ldots, \Pi_M) = \int\rd\rho
{P(\Pi_1, \ldots, \Pi_M|\rho)P(\rho)}. \en
Then, the quantum state
for the remaining systems of the ensemble  becomes  \cite{QBayes} \be \label{E4}
\rho^{(N-M)} =  \int\rd\rho
P(\rho|\Pi_1,\ldots,\Pi_M)\rho^{\otimes N-M}. \en 
Notice that using the quantum de Finetti representation, (\ref{E1}) or (\ref{E4}),  one makes predictions 
also for the correlations between the individual systems of the ensemble. 
As noted  in Ref. \cite{QBayesD}, the exchangeable  representation (\ref{E1})    cannot be carried to the probability
theory formulated in the linear space either over the field of
real or quaternionic numbers, thus being a unique feature of  the complex Hilbert space.  
It was also argued that   an  exchangeable de Finetti state  
is a natural  substitute for the  ``unknown quantum state"  in the QT  \cite{QBayesD} 
and that the latter has to be banished from it. 

 However, as we show below, the way  the quantum de Finetti representation accounts for 
 correlations of the individual systems  of the ensemble bears a considerable problem.  
 The problem lies in the fact that  an exchangeable prior for 
 a quantum state of $N$ exchangeable  systems almost surely assigns  correlations to them, 
 which manifest themselves in nonzero Quantum Discord (QD)  \cite{OllivierZurek}. 
The QD accounts for  non-classical correlations between the individual systems of a composite
system and can be  experimentally measured (see for instance, Ref. \cite{discord}).   
But  as we discuss below,  if it is known that there are no such correlations \cite{Note}  and 
the information on the measured state is limited  (e.g. to   basic symmetries)  one \textit{cannot 
 combine   these two features in a quantum  de Finetti prior.}  This problem is even  worse:  
 the posterior, as given by the exchangeable de Finetti
representation Eq. (\ref{E4}), will also have  a nonzero QD almost
surely (see below).

Let us recall the QD definition \cite{OllivierZurek}. The QD
quantifies  non-classical correlations of  two systems $A$ and $B$ of
a composite system in the Hilbert space
$\mathcal{H}_A\otimes\mathcal{H}_B$. Given a density matrix $\rho$
of a composite state, the QD is the difference between two
versions   of the mutual information, $D_A(\rho) = I(\rho)
-Q_A(\rho)$. One is $I (\rho) = H(\rho_A) + H(\rho_B) - H(\rho)$,
where $H$ is the von Neumann entropy $H(\rho) = -\mathrm{Tr}(\rho
\ln\rho)$ and $\rho_{A,B} = \mathrm{Tr}_{A,B}(\rho)$ are the
reduced density matrices.  The other one is  defined by optimizing
over all possible measurements in $A$ and is given  as follows $Q_A(\rho) = H(\rho_B) -
\mathrm{min}\sum_kp_kH(\rho_{B|k})$, where  $\rho_{B|k} =
\mathrm{Tr}_A(E_k\otimes
\openone_B\rho)/\mathrm{Tr}(E_k\otimes\openone_B\rho)$ is the
state of $B$ conditioned on outcome $k$ in $A$, and  $\{E_k\}$ is
the set of POVM elements. These two formulations of the mutual
information are two quantum generalizations of the classical
mutual information $I(A:B) = (HA) +H(B) - H(A,B)$.  On the other
hand, the state is of zero QD if and only if there exist a von
Neumann POVM $\Pi_k = |\psi_k\rangle\langle \psi_k|$ that
 \be
 \label{E5}
 \sum_k(\Pi_k\otimes\openone_B)\rho(\Pi_k\otimes\openone_B) = \rho,
 \en
i.e. the $\rho$ is a state obtainable  by a von Neumann
measurement, where  only classical correlations remain. The QD  is
currently thought of as a resource for various  classically
intractable tasks including  the quantum computation
\cite{Datta,Cavalcanti,Mahdok}. For instance,  the remote state
preparation  (a variant of the quantum teleportation protocol)
based on the QD only, i.e. without the quantum entanglement, was
already implemented experimentally \cite{Dakic}.

Let us now inspect the QD of the exchangeable state (\ref{E1}).
To this goal we invoke a sufficient condition  \cite{NSQD0} for
non-zero QD   of a bi-partite quantum system,
which is the rank of their correlation
matrix $R_{n,m}$. In our case the said composite consists of two exchangeable systems, i.e. $\rho^{(2)} =
\sum\limits_{n,m}R_{n,m}A_n\otimes B_m$, where $A_n$ and $B_m$ are
bases in the  space of Hermitian matrices acting in $\mathcal{H}_{A,B}$. Then rank$(R) > d_{A}=d_{B}$  implies  $D(\rho)
>0$ (note that for an exchangeable state
the QD is symmetric  with respect to swapping  of systems  $A$
and $B$). Consider the two-dimensional systems, where the calculations
are simplified with the help of  the Bloch
vector representation. In this case, the unique  measure of Eq.
(\ref{E1}) can be cast as $\rd\mu(\rho) =  \rd\mu(\vec{n}) =
\frac{3}{4\pi}\rd n_1\rd n_2 \rd n_3P(\vec{n})$, where the Bloch
vector satisfies $\vec{n}^2\le 1$.  In this case $\rho(\vec{n}) =
\frac12(\openone + \vec{n}\vec{\sigma})$, where
$\vec{\sigma}=\{\sigma_1,\sigma_2,\sigma_3\}$ is the vector of
Pauli matrices. Then,  by  Eq. (\ref{E1}), the
 density matrix $ \rho^{(1)}$ of system $A$ or $B$ is 
\be \label{E6} \rho^{(1)} = \int \rd\mu(\vec{n})\frac12(\openone +
\vec{n}\vec{\sigma}) = \frac12(\openone + \vec{x}\vec{\sigma}),
\en where we have denoted $\vec{x} = \int\rd\mu(\vec{n})\vec{n}$.
Whereas the composite density matrix
reads 
\be \label{E7} \rho^{(2)} =
\frac14\left(\openone\otimes\openone +
\vec{x}\vec{\sigma}\otimes\openone + \openone\otimes
\vec{x}\vec{\sigma} + \sum_{i,j}\tau_{i,j}\sigma_i\otimes\sigma_j
\right) 
\en 
with the matrix $\tau$ defined as $ \tau  =
\int\rd\mu(\vec{n}) \vec{n}\otimes \vec{n}$ (in the tensor product
notation). Now it is a simple observation that the correlation
matrix $R$ of  $\rho^{(2)}$ given by Eq.~(\ref{E7})  in the basis
$(\openone, \sigma_1,\sigma_2,\sigma_3)$ has the following
block-matrix form \be \label{E8}
R = \frac14\left(\begin{matrix}1 & \vec{x}^T \\
\vec{x} & \tau\end{matrix}\right), \en hence if
rank$(\tau)=3>\mathrm{dim}(\mathcal{H_A})=2$ then
D($\rho^{(2)})>0$. But $\tau$ is full rank for    any distribution
$P(\vec{n})$  provided that it has three-dimensional domain of
support in the Bloch ball.   The simplest example of this class  is the
point-mass distribution with $P(\vec{n}) = \sum_\alpha
p_\alpha\delta(\vec{n}-\vec{e}_\alpha)$, where all $p_\alpha>0$
and $\vec{e}_1$, $\vec{e}_2$, and $\vec{e}_3$ being \textit{any}
three  linearly independent Bloch vectors.

Next we evaluate the geometric measure of the QD proposed for the
two-qubit system in Ref. \cite{NSQD0}, which gives the distance to
the zero QD states in the Bloch vector space. It reads
\be \label{E9} D (\rho^{(2)}) = \frac14\left( ||\vec{x}||^2
+||\tau||^2 - \lambda_\mathrm{max} \right), \en where $||\tau||^2
= \mathrm{Tr}(\tau^T\tau)$ (in our case $\tau^T = \tau$) and
$\lambda_\mathrm{max} $ is the maximal eigenvalue of the matrix
$\Lambda= \vec{x}\vec{x}^T+ \tau\tau^T$.  Simple calculations give
\begin{eqnarray}
\label{E10}
D (\rho^{(2)}) &=& \frac14{{}_{\textstyle{\mathrm{min}}}\atop\scriptstyle{\vec{m}_1^2=1}}\Biggl(\int\rd\mu(\vec{n}_1)\int\rd\mu(\vec{n}_2)(1+\vec{n}_1\vec{n}_2)\nonumber\\
&& \times \vec{n}_1\biggl[\sum_{j=2,3}\vec{m}_j\otimes \vec{m}_j\biggr]\vec{n}_2\Biggr),
\end{eqnarray}
where the vectors $\vec{m}_j$, $j=1,2,3$, form an  orthonormal
basis in the Bloch space. Eq. (\ref{E10}) shows that almost surely
$D(\rho^{(2)})>0$, since the
matrix in the square brackets in Eq. (\ref{E10}) is manifestly
positive for almost all choices of the measure $\rd\mu(\vec{n})$,
while  the  scalar factor preceding it is also positive. The only
exception is the case of a measure $\rd\mu(\vec{n})$ which has
support  on the one-dimensional vector space (parallel to some
vector $\vec{m}_1$) i.e. when $P(\vec{n}) $ is \textit{a distribution
confined to a line in the Bloch sphere}. For the same  reason, the nonzero QD
 will almost surely prevail  for any finite number of measurements.   
 This is,  in fact, a quite general feature of the QD, since it was shown 
\cite{QDNever0} that the   states of zero QD belong to a  zero measure subset of all states. 
Thus,  one is led to accept a nonzero QD for the \textit{result}
of the Bayesian reconstruction for any finite number of
measurements, if the de Finetti  exchangeable density matrix is
admitted as a prior.   The Bayesian experimentalist making measurements on one system at a time will not be confused by this, but
if  a more advanced set-up is to be used with joint measurements  on two or more systems at
a time, to measure  their correlations,  the problem  is bound to arise due to the way the exchangeable representation
assigns such  correlations. 

The strength of the Bayesian approach lies in
selecting \textit{a judicious prior}, reflecting \textit{all } the
information available at hand. This  is a recurrent theme of the
Bayesian Statistical Inference in general \cite {JeffreysBook,JaynesBook}. For
instance, if one knows somehow that there are no quantum
correlations between the individual systems  of the ensemble (e.g.
they are created one at a time), one would like to reflect this in
the prior information. But since one does not know the exact state
of the systems in the ensemble, e.g. only some symmetry
considerations are known for the density matrix parameters,  one
is forced to select a straight line in the Bloch ball  for  the
prior distribution to have zero QD. But which one should be
selected? In the current situation the experimentalist has really
no clues for choosing it. A limited prior information on the
actual state of the system (a typical case of the QT) and zero QD taken 
jointly  \textit{do not allow}  one select a measure in representation Eq.
(\ref{E1}) for the state of the ensemble which incorporates all
the available information. \textit{Thus, the exchangeable representation (\ref{E1})   is not an
universal prior that fits all (even the most typical) cases}.

It is to be noted that the described problem 
has rather deep roots in the  linearity of the Quantum Mechanics
itself. The field of Quantum Statistical Inference viewed as a variant of the general 
Parametric Statistical Inference  introduces one special feature: the 
linearity of the probability assignment on the estimated parameter, i.e. the
 density matrix $\rho$ describing the quantum state. 
 This unique feature forces one to mix one's incomplete knowledge on the 
 parameter to be estimated and the parameter itself.  This is seen  already on the single system level. A
prior  with the density $P(\rho)$ for the estimated density matrix  $\rho$  invariably leads  to a new density 
matrix $\rho_{est}  = \int \rd\rho
P(\rho) \rho$ by the total probability assignment
\begin{eqnarray}
\label{E11}
p(\Pi_\alpha) &=&\int\rd\rho P(\rho) p(\Pi_\alpha|\rho) = \int \rd\rho P(\rho) \mathrm{Tr}(\Pi_\alpha\rho) \nonumber\\
&=&  \mathrm{Tr}(\Pi_\alpha \rho_{est}),
\end{eqnarray}
where the passage from the first line to the second is provided by
linearity of Born's rule (and convexity of the set of density
matrices) with the acceptance of the result $\rho_{est}$ as the
``quantum state" by Gleason's theorem (we note in passing that  usage of Gleason's result is an important  implicit step in the
simple and elegant proof of the quantum de Finetti   representation  in
Ref. \cite{QBayesD}).   From this point of view,  the experimentalist in the process of the QT  
extracts the actual quantum state from such a mixture.    
Indeed,  the Bayesian updating in the QT is not an unlimited  process, 
but  has as a limit the maximal possible information 
obtainable from the ensemble (for instance, the Jones limit  \cite{Jones} for 
the   pure state QT with the unitarily invariant prior).  After the maximal possible 
information is extracted (to the inevitable  imperfections of 
experimental apparatus and restriction to finite number of measurements) 
no further update is possible and the experimentalists  concludes what is the \textit{actual} 
state of the ensemble.

In conclusion,  we have shown that the  exchangeable, i.e. the quantum
de Finetti, representation almost surely  assigns a nonzero Quantum Discord to ensemble of 
exchangeable systems.  If  preparation of uncorrelated
systems is assumed, the   simultaneous
requirements of zero Quantum Discord and exchangeability of the
prepared ensemble of quantum states  do not allow selection of any
prior at all within the quantum de Finetti representation.
Furthermore, we point out that it is the linearity of  the Born
rule for the probability assignments in Quantum Mechanics that
leads to such contradictory requirements since, by Gleason's theorem, one mixes  
one's knowledge about the state and the
representative of the state in one density matrix.

 This work was supported by by Belarusian agency
BRRFI (D.M.), and also by the FAPESP and CNPq grant agencies of
Brazil.

\newpage
\section{Supplementing Material: Review Reports and Replies}

In this section we reprint the Referee reports which, in our opinion, 
add to the points considered in the manuscript.

\subsection{J. Phys. A: Math.    Theor.  }

\subsubsection{Board Member Report}

This paper makes the (correct) observation that states of de Finetti
form
  generally have nonvanishing quantum discord. I believe this is a
  novel
  and
  interesting observation. It is not a deep result, however. Given
  the
  tools
  provided by previous authors, the proof consists simply of checking
  the
  definition, in a few lines of simple algebra.
 
  Although the authors refer centrally to the quantum de Finetti
  theorem,
  their
  result has nothing to do with this theorem. States of de Finetti
  form
  (their
  equation 1) occur naturally in the analysis of quantum state
  tomography.
 
  I believe the authors are mistaken in their evaluation of the
  relevance of
  their result. They conclude that the latter implies that states of
  de
  Finetti
  form should not generally be used as priors in quantum tomography.
  That
  conclusion is simply not warranted and not implied by their
  argument.
 
  For these reasons I cannot recommend publication of this
  manuscript.
 
 \subsubsection{Referee Report}
  
  The main result is that under the assumption than one has an
  uncorrelated   quantum system, considering infinitely extendable exchangeable
  systems, the   result of using the quantum Finetti representation for a Bayesian
  Quantum   Tomography will be a correlated quantum state.
 
  This is an interesting and clearly written paper. The problem is
  clearly  introduced, the relevant references are present in the
  bibliography,   its   length   is appropriate and the results seem correct. My main concern is
  with   one of   the   main assumptions made by the authors, that the state is initially
  uncorrelated.   From my point of view this is a big assumption.
 
  It is completely fine to assume what would happen if one had, in
  theory, an   uncorrelated initial state. But in this paper the authors seem
  concerned   with   what should do an experimentalist in this case to reconstruct the
  state of   the   system without introducing additional correlations. I think that it
  is   really   difficult to assume that in a real experiment one is working with a
  completely   uncorrelated system. Any small perturbation, which is inevitable
  when   performing a measurement, would perturb the systems in such a way
  that   it   becomes correlated. Even if one assumes that the different parts of
  the   systems   are prepared independently in distant laboratories, the measurement
  process   which is essential for the tomography process would perturb the
  system.   This is   clearly explained in Ref. [31].
 
  The authors closely follow Ref. [12] when discussing the role of
  the   quantum   Finetti representation in Bayesian Quantum Tomography. According to
  that   paper,   just after Eq. (1.5): "quantum-state tomography is not about
  uncovering   some  ''unknown state of nature'', but rather about the various
  observers'   coming   to   agreement over future probabilistic predictions". If one agrees
  with   this   statement, and also accepts the result from Ref. [31] about how
  difficult   is it   to actually have an uncorrelated state, shouldn't we expect that
  the   result   of   a Bayesian quantum tomography process is actually a correlated
  quantum   state?   That would be the only way to predict the effects of the
  correlations   that   one   would introduce in the system while manipulating it.
 
  In summary, I think that this is an interesting paper and that it
  deserves   being published in some form, but I would advise the authors to
  support   it's   main assumptions (an initial uncorrelated system and the need to
  obtain an   uncorrelated state) with stronger arguments. I also think that,
  unless   the   authors can proof that the correlations obtained through this
  quantum   Bayesian   tomography are large, the results obtained in this work are what
  one   could   expect from  the  results in Ref. [31]. For this reason it is my
  opinion   that    this work is not novel enough to be published as a fast track
  communication. I   would be inclined to recommend it to be published as a regular
  article   if   the   authors could provide additional explanations for the points raised
  before.  
 
\subsection{Phys. Rev. A}
There were three rounds of review.

\subsubsection{Round 1}

The manuscript "The quantum de Finetti representation for the Bayesian 
Quantum Tomography and the Quantum Discord" is a short note on how 
badly the three topics mentioned in the title play together. The idea, 
as I understood it, is the following: The de Finetti representation is 
a symmetrization over different realizations of a quantum state. In a
Bayesian scenario, one eventually assigns a probability distribution 
over the symmetrized quantum state. This state will naturally belong 
to the class of symmetric separable states. However the symmetric 
states with zero quantum discord are only a set of measure zero among 
them. This seems to bother the authors. 
 
I cannot follow the arguments of the authors. The most obvious reason
is that if the problem occurs for classical states, then it will in
particular appear for product states. Second, there is no conflict at
all: The symmetrized Bayesian estimate is not used to make prediction
about the correlations between the individual copies (this does
already not work in the classical de Finetti scenario), but rather
about single instances. Finally, in a Bayesian scenario one aims to
minimize the cost for a (erroneous) future prediction. Since the set
of classical states is of measure zero, it would be extremely risky
bet for a classical state. Therefore a Bayesian procedure is actually
not expected to yield such a result, contrary to what the authors
implicitly claim.

The issues mentioned above already lead to my clear recommendation
against a publication.

\medskip
\textit{Reply to Referee 1}
\medskip

This referee recommendation for rejection is based on outright misleading statements, some of 
them are even wrong, as can be verified in the graduate textbooks  on probability theory. 

The referee is also expected to read the manuscript to the very end before 
judging on its validity and it does not seem to have occurred in this particular case.

Our manuscript touches on the recurrent issue of the Quantum Mechanics: the reality of the quantum state. This issue is of continuous debate to this very day.
The quantum de Finetti representation for the Quantum Tomography was proposed to replace the notion of an unknown state from the Bayesian Quantum Tomography by  making a claim about the  universality of this prior.  We  summarize their claim: it \textit{was} proposed  as a universal prior for  the \textit{whole ensemble} of the exchangeable states (which can be consulted in Refs. [14,15]).  

The classical de Finetti representation involves \textit{joint} probability of an infinite sequence of exchangeable variables, \textit{quite contrary} to the referee statement.  

We show that if one takes the quantum de Finetti representation as an universal prior -- i.e. seriously --  one immediately comes to a contradiction. We show that the roots of this contradiction are in the linearity of the Quantum Mechanics, which makes one to mix the knowledge on the state with the state itself -- a procedure not possible in the general classical statistics where the Bayesian probability and the unknown parameters are not mixed in one object.

\subsubsection{Round 2}

A main finding of this paper is that almost all exchangeable priors
over a collection of subsystems give rise to non-zero discord, and
hence non-classical correlations between the subsystems. It is then
argued that this is problematic in a situation where it is "known"
that there are no such correlations. The authors suggest that,
therefore, one cannot reasonably represent the fact that there are no
non-classical correlations by a de Finetti prior. They then argue that
this is due to the fact that in quantum theory "one mixes one's
knowledge about the state and the representative of the state in one
density matrix."

While I agree with the very last sentence, I think it is also the
source of a confusion that led the authors to draw wrong conclusions.
Take, as an example, an experiment where one player (let us call him
A) prepares a string of identical bits, e.g., 000000. There are
obviously no correlations in this bit string. However, another player
(B) may only know that Player A prepares strings consisting of
(uncorrelated) identical bits, but does not know whether it is 000000
or 111111. If we now describe the knowledge of B by a probability
distribution (e.g., one that assigns probability 1/2 to 000000 and 1/2
to 111111) then the subsystems are clearly strongly correlated. But
this correlation is simply due to B's ignorance of one bit (rather
than a "physical" correlation generated by Player A). Note that this
issue has been extensively discussed in the literature, see, e.g.,
Phys. Rev. Lett. 109, 120403 (2012).

Summarizing, I agree with the authors' technical claim that there is
non-zero discord in the considered de Finetti states. However, I do
not see any reason why this should be physically relevant or why one
should be concerned about this. It may well be that the authors have a
point that I (as well as the other referee) missed, but in this case
it may be worth if they rewrite their manuscript and try to make their
line of argument clearer.

In view of the above assessment, I recommend not to accept this paper
for publication.

\medskip 
\textit{Reply to Referee 2}
\medskip

We have carefully analyzed what the second Referee has written in the objections part of his/her report and have found that:
(i) There is no \textit{concrete} statement of what is actually wrong in our arguments. His/her statement "While I agree with the very last sentence, 
I think it is also the source of a confusion that led the authors to draw wrong conclusions." \textit{does not } point what conclusions are wrong and why. 
Moreover, the Referee agrees with \textit{all} technical part of the manuscript.  

(ii) His/her example of spurious correlations is both \textit{wrong} and \textit{irrelevant}. First of all, we consider the quantum correlations which  are not possible in the classical setting.  We do not consider any classical correlations contained in the density matrix.
Second, we are surprised to find out that his/her knowledge of the elementary probability theory is deficient. Indeed, let us cite the report.

\textbf{Referee:} Take, as an example, an experiment where one player (let us call him A) prepares a string of identical bits, e.g., 000000. There are
obviously no correlations in this bit string.

\textbf{Our remark:} The string "00000" , i.e. $x_1 x_2 x_3$, etc, where all $x_k = 0$, $k =1,2,3 ...$,  consist of the maximally correlated  random  values $x_k$ (i.e. result of a particular measurement is perfectly correlated with the number of this measurement).

\textbf{Referee:}  However, another player (B) may only know that Player A prepares strings consisting of
(uncorrelated) identical bits, but does not know whether it is 000000 or 111111. If we now describe the knowledge of B by a probability
distribution (e.g., one that assigns probability 1/2 to 000000 and 1/2 to 111111) then the subsystems are clearly strongly correlated. But
this correlation is simply due to B's ignorance of one bit (rather than a "physical" correlation generated by Player A). 

\textbf{Our remark:}   This example is wrong.  In the present formulation, there are correlations in both cases, in either version of the description by  player A or by player  B.   Both players believe and would agree upon   the same thing -- correlations.  

Finally, this is \textit{not} what we consider in the quantum case.  In the quantum case player A has a tensor product of equal density matrices ($\rho$). His systems are \textit{not} quantum correlated (trivially being only \textit{classically} correlated!).  Player B  on the other hand, if he assumes the quantum de Finetti representation, would have \textit{quantum} correlated individual systems. given by a nonzero quantum discord (see our manuscript).  Obviously, this is \textit{quite contrary}  to the classical case presented by the Referee. Why the Referee has brought up this example is totally mysterious for us.   Moreover, the quantum correlations lead to predictions of observable events, quite independent of the local observer or his/her beliefs, and thus must be taken seriously.

There is also  suggestion of an irrelevant reference  in his/her report:

\textbf{Referee:} Note that this issue has been extensively discussed in the literature, see, e.g.,
Phys. Rev. Lett. 109, 120403 (2012).

\textbf{Our remark:}  While being an important result, it hardly touches  the issues raised in the manuscript.

\textbf{Referee:} Summarizing, I agree with the authors' technical claim that there is  non-zero discord in the considered de Finetti states. However, I do
not see any reason why this should be physically relevant or why one should be concerned about this.

\textbf{Our reply:}    Thus,  the Referee  makes the following point: it is of \textit{no importance}  if there are any spurious quantum  correlations in the prior used for Bayesian tomography.  We remind that the Bayesian approach is in its essence the updating procedure on the "knowledge" (we follow the Referee style  and  put this word in quotes).  We then observe the following: if there are the quantum discord correlations, they are passed with a high probability to the posterior state (see our manuscript). The correlations are measurable and lead to observable essentially quantum effects (see in the manuscript for the references).  Thus the observer consistently  fools him/herself, which can de detected by a joint measurement on the rest of the  systems. 

The problem is not the Bayesian methods, of course.  It is the only possible extension of the usual mathematical logic including the probabilistic reasoning (see for instance, the last of our references, the excellent book by late E. T. Jaynes).  The problem is in the prior, which does not allow one to account for absence of the quantum correlations which is implied by the very design of the \textit{most} tomographic  experiments (in most of the cases the systems are prepared in different time slots and are not quantum correlated).   Why then the Referee credits his/her "knowledge", that  the systems are exchangeable, and discards the "knowledge" of absence of the quantum discord correlations?  More importantly, where his/her "knowledge" of the exchangeability come from?     We remark that  we can prove or disprove our "knowledge" of absence of any quantum correlations  by making use of the user/observer independent  quantum resource -- the presence of the quantum discord, whereas he/she can hardly  to do the same with his/her "knowledge". Who's prior assumptions are then more correct?  We believe the answer is obvious. 

We conclude that while the Referee has agreed to all the technical part of the manuscript, he/she has also found \textit{no} errors in our arguments, by \textit{not} providing any concrete example of such. His/her example, on the basis of which he/she seem to reject our conclusions is both wrong (from the point of view of the basic probability theory) and irrelevant. We do not understand why the Referee has such conclusion. We ask for  another round of the review and urge for a thoughtful reading of the manuscript by future Referees. 

\subsubsection{Round 3}

Quantum Bayesian inference as introduced by Jones and later on
discussed and studied by several authors (see references in the
manuscript) is used to estimate (classical) information encoded in a
quantum system. Specifically, one can assume that information about
the state of a single qubit is encoded (via a preparation procedure)
into a single two-level system (physical representation of a qubit).
To be more explicit about the encoding stage: a specific reference
(known) state of a physical qubit (let say prepared in the state
$\rho_0$) is unitarily rotated (transformed). The two angles
associated with such SU(2) rotation represent the classical
information encoded into a single physical system.

Later on this information about the two angles (a qubit state) can be
recovered by performing a measurement on the single two-level system.
If the measurement is optimal [in a sense as described by Holevo, or
Massar and Popescu, and Derka et al.] then using a prior knowledge
about the situation one can make an estimation of the information that
has been encoded into the quantum system. There exists a strict bound
on the fidelity of such estimation. It is well known that the fidelity
of estimation is equal to 2/3 providing it is assumed that the angles
of the SU(2) transformation have been encoded via rotation of a pure
state $\rho_0$.

However, the information about the single-qubit state can be encoded
into a finite ensemble of physical qubits. These qubits can be
initially prepared in a product state $\rho^{\otimes N}$ or more complex reference state, e.g. entangled state of $N$ physical
qubits. These physical qubits are then rotated in order to encode an
information about a single qubit [different representations of SU(2)].
Given different encoding schemes (of the same information about a 
state of a single qubit!) one uses different measurements in order to
optimally recover the encoded information. For instance, if the
information about a single qubit is encoded in two identically rotated
physical qubits then one can perform on these two qubits either local
measurements or non-local measurements. As shown by Massar and Popescu
and Derka et al. optimal non-local measurements lead to a better
estimation of the information about the single qubit than that when
the information was encoded into two identically prepared physical
qubits. This well known result can be extended to the case when the
SU(2) rotation (information about a single-qubit) is encoded into $N$
physical qubits. If the initial the reference state of $N$ physical
qubits is a pure state $\vert 0\rangle^{\otimes N}$ then the fidelity
of information about the single-qubit state encoded in this system can
be recovered with the fidelity $(N+1)/(N+2)$. It has to be stressed
that the information about the single-qubit state can be encoded into
highly-entangled physical qubits systems. In fact one can find an
optimal $N$-qubit (entangled) state into which an information about a
single qubit is encoded optimally (see papers by Gisin et al., Bagan
et al. and also by Rapcan et al. PRA 84, 032326 (2011)) and can also
be recovered optimally.

This is a typical task of quantum Baysian inference of information
about a single-qubit state. Obviously, one can ask a question about an
(optimal) encoding and decoding of two-qubit state into a pair of
physical qubits or a bigger ensemble of physical qubits. Here the task
will be to use the procedure described above to estimate a two-qubit
state. A two-qubit state at the stage of encoding can be entangled,
but if incomplete measurement is performed the estimated two-qubit
state might not exhibit any entanglement. But opposite can't happen
providing one is using the Baysian inference properly, that is, if the
information encoded into quantum systems corresponds to non-entangled
(separable) two-qubit state, the reconstructed two-qubit state can't
exhibit entanglement. And this is irrespective whether the information
is encoded into two physical qubits or a large ensemble of physical
qubits.

I am writing this (rather trivial comment) in order to show that the
problem discussed by the authors in their paper is artificial. I
believe the first referee made it very clear when s/he wrote: The
symmetrized Bayesian estimate is not used to make prediction about
correlations between individual copies but rather about single
instances. That is, the authors are using the single-qubit
estimation procedures (as described in my comments above) and they are
trying to make conclusions about correlations between qubits in the
ensemble.

Given the character of responses to reports of previous referees I
might expect that the authors will strongly disagree with my
conclusion, but in my humble opinion the paper is not suitable for
publication since it deals with an artificial problem that does not 
exist providing the quantum Bayesian estimation is correctly applied.


\begin{thebibliography}{99}
\bibitem{Note} For example,   the systems may be prepared in the same state one by one sequentially.
\end{thebibliography}
\end{document}